\documentclass{ws-ijmpa}

\begin{document}

\markboth{A. Bazavov, B.A. Berg, A. Velytsky} {Time Evolution of
the Structure Factors}

%
\catchline{}{}{}{}{}
%

\title{EVOLUTION OF THE STRUCTURE FACTORS IN PURE SU(N) LATTICE
GAUGE THEORY AND EFFECTIVE SPIN MODELS}

\author{\footnotesize ALEXEI BAZAVOV, BERND A. BERG\\and ALEXANDER VELYTSKY
\footnote{Present address: Department of Physics and Astronomy,
University of California, Los Angeles, CA 90095-1547, USA } }

\address{Department of Physics, Florida State University\\
         Tallahassee, FL~32306-4350, USA\\
         School of Computational Science, Florida State University\\
         Tallahassee, FL~32306-4120, USA}


\maketitle

\pub{Received 1 November 2004}{}

\begin{abstract}
We consider model~A dynamics for a quench from the disordered into
the ordered phase of SU(3) lattice gauge theory and the analogue
$3d$ 3-state Potts model. For the gauge model this corresponds to
a rapid heating from the confined to the deconfined phase. The
exponential growth factors of low-lying structure function modes
are numerically calculated. The linear theory of spinodal
decomposition is used to determine the critical modes. This allows
for the Debye screening mass estimation in an effective
phenomenological model. The quench leads to competing vacuum
domains, which make the equilibration of the QCD vacuum after the
heating non-trivial. The influence of such domains on the gluonic
energy density is studied.

\end{abstract}

\section{Introduction}

In Ref.~\refcite{BHMV,BMV} it is emphasized that in studies of the
QCD deconfining transition (or cross-over) by means of heavy ion
experiments, one ought to be concerned about non-equilibrium
effects after a rapid {\it heating} of the system. This is because
the heating may be modelled as a rapid quench and the QCD high
temperature vacuum is characterized by ordered Polyakov loops
which are similar to the low temperature phase of analogue spin
models. The subsequent evolution leads to vacuum domains of
distinct $Z_3$ triality, and one ought to be concerned about
non-equilibrium effects.

Here we extend the investigation to SU(3) lattice gauge theory. We
report preliminary results about the influence of such domains on
the gluonic energy density and pressure of pure SU(3) lattice
gauge theory.

The Markov chain Monte Carlo (MC) process provides model~A
(Glauber) dynamics in the classification of Ref.~\refcite{ChLu97}.
As a time step a sweep of systematic updating with the
Cabibbo-Marinari \cite{CaMa82} heat-bath algorithm and its
improvements of Ref.~\refcite{FaHa84,KePe85} is used (no
over-relaxation, to stay in the universality class of Glauber
dynamics). Although this is certainly not the physical dynamics of
QCD, in the present state of affairs it appears important to
collect qualitative ideas about eventual dynamical effects. For
this purpose the investigation of any dynamics, which actually
allows for its study ought to be useful.

\section{Preliminaries}

\subsection{Two-Point Correlation Function}

Consider two-point correlation functions defined by
\begin{equation}\label{e:L0LR_def}
    \langle u_0(0)u_0^\dagger(\vec{j})\rangle_L=\frac{1}{N_\sigma^3}
    \sum_{\vec{i}}u_0(\vec{i})u_0^\dagger(\vec{i}+\vec{j}),
\end{equation}
where $u_0(\vec{i})$ is the relevant fluctuation about some average
and $\vec{i}$ denotes spatial coordinates. Periodic boundary conditions
are used and the subscript $L$ on the left-hand side reminds us that
the average is taken over the spatial lattice.
For gauge systems we deal with fluctuations of the Polyakov loop,
and for analogue spin systems with fluctuations of the magnetization.

The finite volume continuum limit of (\ref{e:L0LR_def}) is achieved by
lattice spacing $a\rightarrow0$, $N_\sigma\rightarrow\infty$ with the
physical length of the box $L=aN_\sigma=const$. This means that
\begin{equation}\label{e:L0LR_contlim}
    \langle u_0(0)u_0^\dagger(\vec{j})\rangle_L=\frac{1}{a^3N_\sigma^3}
    \sum_{\vec{i}}a^3 u_0(\vec{i})u_0^\dagger(\vec{i}+\vec{j})
\end{equation}
transforms into
\begin{equation}\label{e:LOLR_contlim2}
    \langle u(0)u^\dagger(\vec{R})\rangle_L=\frac{1}{L^3}
    \int d^3r\,u(\vec{r})u^\dagger(\vec{r}+\vec{R}),
\end{equation}
with $\vec{r}=a\vec{i}$, $\vec{R}=a\vec{j}$,
$u(\vec{r})=u_0(\vec{i})$, and so on.

\subsection{Structure Functions}

We define structure function as Fourier transform of the
two-point correlation function (\ref{e:LOLR_contlim2}):
\begin{equation}\label{e:SF2_def}
    F(\vec{p})=\int \langle u(0)u^\dagger(\vec{R})\rangle_L\,
    e^{i\,\vec{p}\,\vec{R}}\, d^3R.
\end{equation}
Occasionally it is also convenient to use structure functions
normalized by the volume:
\begin{equation}\label{e:SF1_def}
    S(\vec{p})=\frac{1}{L^3}\int \langle u(0)u^\dagger(\vec{R})
    \rangle_L\, e^{i\,\vec{p}\,\vec{R}}\, d^3R.
\end{equation}
Periodic boundary conditions imply:
\begin{equation}\label{e:k_def}
    \vec{p}=\frac{2\pi}{L}\vec{n},
\end{equation}
where $\vec{n}$ is an integer vector $(0,0,0)$, $(0,0,1)$, etc.

The discretized version of (\ref{e:SF2_def}) is
\begin{equation}\label{e:SF2_disc}
    F(\vec{p})=\sum_{\vec{j}}a^3\,\langle u_0(0)u_0^\dagger(\vec{j})
    \rangle_L\, e^{i\,a\,\vec{p}\,\vec{j}}.
\end{equation}
Using definition (\ref{e:L0LR_def}) and shifting summation on $\vec{j}$
one arrives (after simple algebra) at the expression
\begin{equation}\label{e:SF2_disc4}
   F(\vec{p})= \frac{a^3}{N_\sigma^3}\left|\, \sum_{\vec{i}}
   e^{-i\,a\,\vec{p}\,\vec{i}}\, u_0(\,\vec{i}\,) \,\right|^2.
\end{equation}
Rewriting the product in the exponent as
\begin{equation}\label{e:aki}
    a\,\vec{p}\,\vec{i}\,=\,a\,\frac{2\pi}{L}\,\vec{n}\,\vec{i}\,=\,
    a\,\frac{2\pi}{aN_\sigma}\,\vec{n}\,\vec{i}\,=\,
    \frac{2\pi}{N_\sigma}\,\vec{n}\,\vec{i}.
\end{equation}
and using definition (\ref{e:SF1_def}) we shape $S(\vec{p})$ into the
form used in our simulations:
\begin{equation}\label{e:SF1_disc2}
    S(\vec{p})= \left|\, \frac{1}{N_\sigma^3}\sum_{\vec{i}}
    \exp\left\{-\frac{2\pi i}{N_\sigma}\,\vec{n}\,\vec{i}\right\}
    u_0(\,\vec{i}\,) \,\right|^2.
\end{equation}
As we let system evolve with time quantity $u_0(\vec{i})$ becomes
time-dependent: $u_0(\vec{i},t)$. The time $t$ corresponds to the
dynamical process, i.e., in our case the Markov chain MC time. We
consider an ensemble of systems (replica) and dynamical observables are
calculated as ensemble averages denoted as $\langle...\rangle$.
Then time-dependent structure functions averaged over replicas are:
\begin{equation}\label{ens_aver_S}
    \hat{S}(\vec{p},t)=\left\langle S(\vec{p},t)\right\rangle=
    \left\langle\left|\frac{1}{N_\sigma^3}\sum_{\vec{i}}
    \exp\left\{-\frac{2\pi i}{N_\sigma}\,\vec{n}\,\vec{i}\right\}
    u_0(\vec{i},t)\right|^2\right\rangle.
\end{equation}
Similarly,
\begin{equation}\label{ens_aver_F}
    \hat{F}(\vec{p},t)=L^3\hat{S}(\vec{p},t).
\end{equation}
Early time evolution of the structure functions after the quench
is governed by the linear theory. The linear approximation results
in the following equation for the structure function:
\begin{equation}
  \frac{\partial \hat{S}(\vec{p},t)}{\partial t}
  = 2\,\omega(\vec{p})\,\hat{S}(\vec{p},t)\,,
\end{equation}
with the solution
\begin{equation} \label{str_fact}
  \hat{S}(\vec{p},t) \,=\, \hat{S}(\vec{p},t=0)\exp\left(
  2\omega(\vec{p})t\right)\,,\qquad \omega(\vec{p}) > 0
  ~~{\rm for}~~|\vec{p}|>p_c\,,
\end{equation}
where $p_c>0$ is a critical momentum. Originally the linear theory was
developed for model~B \cite{CaHi58,Ca68}. Details for model~A can be
found in Ref.~\refcite{BMV}.

During our simulations the structure functions are averaged over
rotationally equivalent momenta and the notation $\hat S_{n_i}$ is
used for the structure function at momentum
\begin{equation} \label{momenta}
  \vec{p} = {2\pi\over L}\,\vec{n}~~~{\rm where}~~~|\vec{n}|=n_i\,.
\end{equation}
The $\hat S_{n_i}$ are called structure function modes or
structure factors. We recorded the modes (including the
permutations) $n_1$: $(1,0,0)$, $n_2$: $(1,1,0)$, $n_3$:
$(1,1,1)$, $n_4$: $(2,0,0)$, $n_5$: $(2,1,0)$, $n_6$: $(2,1,1)$,
$n_7$: $(2,2,0)$, $n_8$: $(2,2,1)$ and $(3,0,0)$, $n_9$:
$(3,1,0)$, $n_{10}$: $(3,1,1)$, $n_{11}$: $(2,2,2)$, $n_{12}$:
$(3,2,0)$, $n_{13}$: $(3,2,1)$, $n_{14}$: $(3,2,2)$, $n_{15}$:
$(3,3,0)$, $n_{16}$: $(3,3,1)$, $n_{17}$: $(3,3,2)$, $n_{18}$:
$(3,3,3)$. Note that there is an accidental degeneracy in length
for $n_8$.

\section{Numerical Results\label{sec_numerical}}

\subsection{Structure Function Modes\label{Qsec_sf}}

\begin{figure}
\centerline{\psfig{file=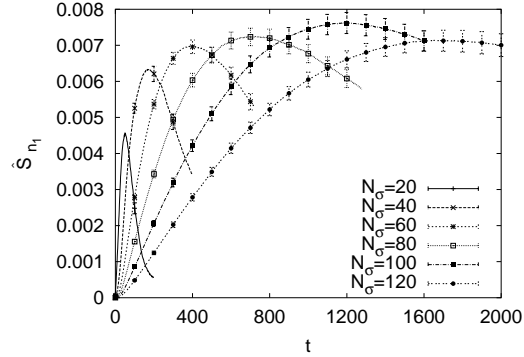,width=7cm}} \vspace*{8pt}
\caption{The first structure function mode for the 3D 3-state Potts
model on $N_\sigma^3$ lattices.} \label{fig_potts_sf01}
\end{figure}

\begin{figure}
\centerline{\psfig{file=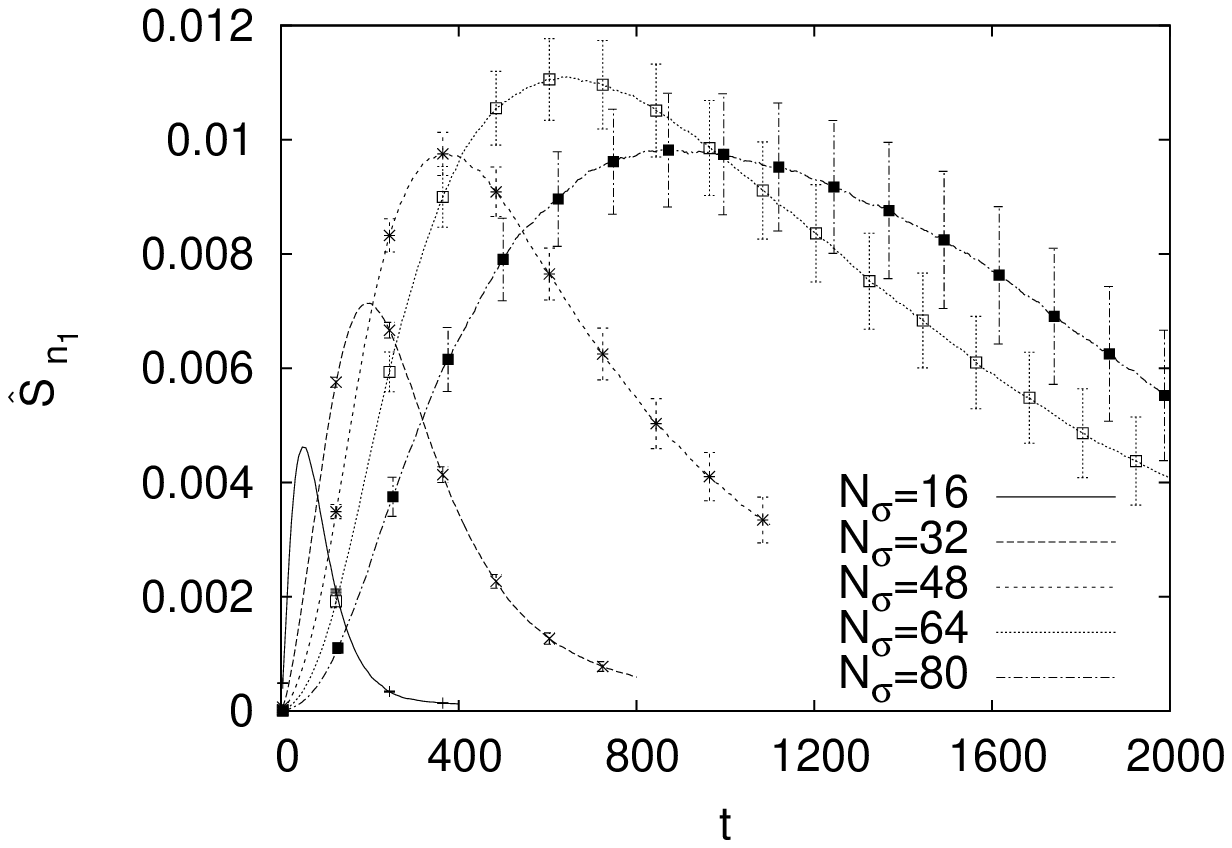,width=7cm}}
\vspace*{8pt} \caption{The first structure function mode for pure
SU(2) lattice gauge theory on $4\times N_{\sigma}^3$ lattices.}
 \label{fig_sf01_su2}
\end{figure}

\begin{figure}
\centerline{\psfig{file=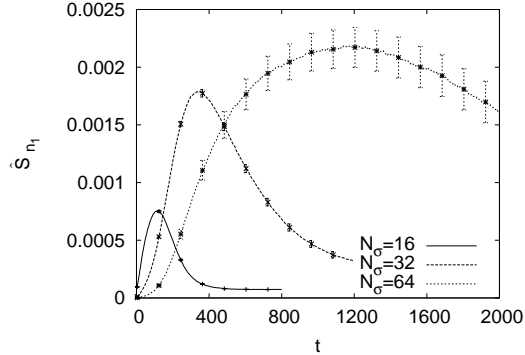,width=7cm}}
\vspace*{8pt} \caption{The first structure function mode for pure
SU(3) lattice gauge theory on $4\times N_{\sigma}^3$ lattices.}
 \label{fig_sf01}
\end{figure}

In Figs.~\ref{fig_potts_sf01}, \ref{fig_sf01_su2}, \ref{fig_sf01}
the time evolution of the first structure function mode $\hat S_{n_1}$
after a heating quench is depicted (from $\beta = 0.2\to 0.3$ in the
3D 3-state Potts model, from $\beta = 2.0\to 3.0$ in pure SU(2) and
from $\beta = 5.5\to 5.92$ in pure SU(3) lattice gauge theory).
Notable in these figures is the strong increase of the maxima
$\hat S_{n_1}^{\max}$ with lattice size. In our normalization of
$\hat S_{n_1}$ non-critical behavior corresponds to a fall-off
$\sim 1/N_{\sigma}^3$ and a second order phase transition to a
slower fall-off $\sim 1/N_{\sigma}^x$ with $0<x<3$. As the
$N_{\sigma}\to\infty$ limit is bounded by a constant, our figures
show that with our lattice sizes the asymptotic behavior may have
been reached for the 3D 3-state Potts model and for SU(2), but not
yet for SU(3) lattice gauge theory.

Our Potts results are from Ref. \refcite{BMV}.
Our results for SU(2) are averages over 10,000 repetitions for
the $4\times 16^3$ lattice, 4,000 for $4\times 32^3$, 800 for
$4\times 48^3$, 340 for $4\times 64^3$ and 106 for the $4\times 80^3$
lattice. For SU(3) we rely on 10,000 repetitions of the quench for
the $4\times 16^3$ lattice, 4,000 for $4\times 32^3$ and 170 for the
$4\times 64^3$ lattice.

\begin{figure}
\centerline{\psfig{file=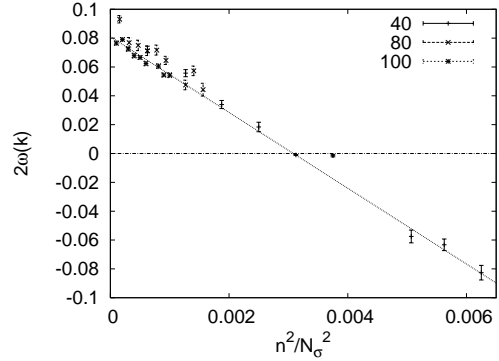,width=7cm}}
\vspace*{8pt} \caption{Determination of $k_c$ for the 3D 3-state
Potts model on $N_\sigma^3$ lattices.} \label{fig_omega_3dsq}
\end{figure}

The results \cite{BMV} of $\omega (k)$ fits from a $\beta =0.2\to 0.3$
quench at zero external magnetic field in the 3D 3-state Potts model are
compiled in Fig.~\ref{fig_omega_3dsq} versus $n^2/N_\sigma^2$.
Approximately, we find straight lines $\omega(k)=a_0+a_1\,n^2$
with a negative slope $a_1$ and we determine the critical momentum
$k_c=2\pi\,n_c/N_\sigma$ as the value where $\omega(k)$ changes
its sign. We find $k_c\approx 0.349$ from our combined data taken
on the $40^3$, $80^3$ and $100^3$ lattices \cite{Vel_thesis}.

\begin{figure}
\centerline{\psfig{file=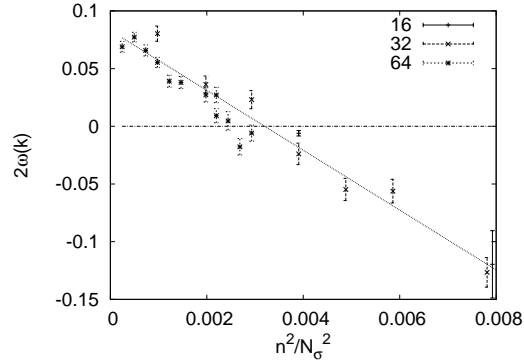,width=7cm}}
\vspace*{8pt} \caption{Determination of $k_c$ for the pure SU(3)
lattice gauge theory on $4\times N_{\sigma}^3$ lattices.}
 \label{fig_su3_omega}
\end{figure}

We use the same techniques to determine the critical mode $p_c$ in
pure SU(3) lattice gauge theory. The results are shown in
Fig.~\ref{fig_su3_omega} and indicate
$a\,p_c=k_c=2\pi\,n_c/N_{\sigma}\approx 0.34$ (from the figure
$n_c^2/N_{\sigma}^2\approx 0.003$). Relying on a phenomenological
analysis by Miller and Ogilvie~\cite{MiOg02}, $p_c$ is related by
$m_D=\sqrt{3}\,p_c $ to the Debye screening mass at the final
temperature $T_f$ after the quench. The relation $k_c/T_f=N_{\tau}\,
a\,k_c$ determines $T_f$ and allows us to convert $m_D$ to physical
units. For our quench we have $T_f/T_c=1.57$ and get
\begin{equation} \label{Debye}
m_D = \sqrt{3}\, N_{\tau}\, a\,k_c\,T_f = 3.7\, T_c\ .
\end{equation}
For pure SU(3) lattice gauge theory $T_c = 265\,(1)\,$MeV holds,
assuming $\sigma=420\,$MeV for the string tension, while for QCD
the cross-over temperature appears to be around $T_c\approx
165\,$MeV, see Ref.~\refcite{Pe04} for a recent review.

In Fig.~\ref{fig_sf01} we observe that not only the heights of the
peaks increases with the spatial volume, but also the time $t_{\max}$,
$\hat S^{\max}_{n_1}=\hat S_{n_1}(t_{\max})$, which it takes to
reach them. Whereas $\hat S^{\max}_{n_1}$ has finally to approach
a constant value, $t_{\max}$ is expected to diverge with lattice
size due to the competition of vacuum domain of distinct $Z_3$
triality.

\begin{figure}
\centerline{\psfig{file=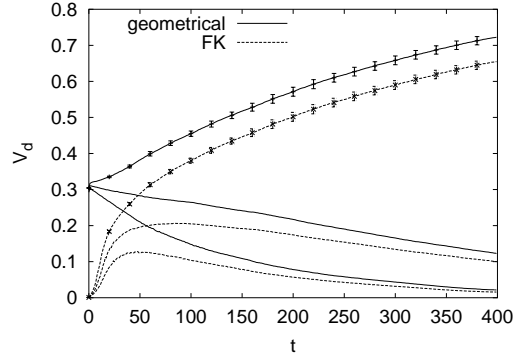,width=7cm}}
\vspace*{8pt} \caption{Largest geometrical and FK clusters for the
3D 3-state Potts model quenched from $\beta=0.2$ to $\beta_f=0.3$
on a $40^3$ lattice.} \label{fig_qnch_3d_geom3d}
\end{figure}

For the Potts models the Fortuin-Kasteleyn (FK) cluster definition
can be used to exactly remap the phase transition into a
percolation model. In case of the 3D 3-state Potts model the
states substitute for the $Z_3$ trialities of SU(3) lattice gauge
theory. In the cluster language the competition of distinct vacuum
domains can be made visible~\cite{BMV}. In
Fig.~\ref{fig_qnch_3d_geom3d} we compare the evolution of
geometrical and FK clusters for a quench of the 3D 3-state Potts
model from its disordered into its ordered phase. We plot the
evolution of the largest clusters for the three Potts
magnetizations in zero external magnetic field $h$. While the
system grows competing FK clusters of each magnetization before
one becomes dominant, geometrical clusters do not compete. This
picture is unfavorable for the use of geometrical clusters of
Polyakov loops in gauge theories, for which the FK definition does
not exist.

The process of competitions between the largest FK  clusters of
different magnetization leads for the proper transition ($h=0$) to
a divergence of the equilibration time in the limit of infinite
systems, an effect known in condensed matter
physics~\cite{ChLu97}. Potts studies \cite{BMV} with an external
magnetic field show that a major slowing down effect survives when
$h\ne 0$ is sufficiently small. As the influence of an external
magnetic field on the Potts model is similar to that of quarks on
SU(3) gauge theory this indicates that the effect may be of
relevance for QCD studies of the crossover region.

\subsection{Gluonic Energy Density}

Although a satisfactory cluster definition does not exist for
gauge theories, the underlying mechanism of competing vacuum
domains is expected to be similar as in the spin models. To study
its influence on the gluonic energy $\epsilon$ and pressure $p$
densities, we calculate these quantities at times $t\le t_{\max}$.

The equilibrium procedure is summarized in
Ref.~\refcite{BoEn96,EnKa00} (in earlier work \cite{De89,EnFi90}
the pressure exhibited a non-physical behavior after the
deconfining transition and the energy density approached the ideal
gas limit too quickly because the anisotropy coefficients were
calculated perturbatively). We denote expectation values of
space-like plaquettes by $P_\sigma$ and those involving one time
link by $P_\tau$. The energy density and pressure can then be cast
into the form
\begin{equation}\label{e:eplusp}
  (\epsilon+p)/T^4 =8N_cN_\tau^4g^{-2}\left[1-\frac{g^2}{2}
  [c_\sigma(a)-c_\tau(a)]\right] (P_\sigma-P_\tau)
\end{equation}
and
\begin{equation}\label{e:eminus3p}
  (\epsilon-3p)/T^4 =12N_cN_\tau^4\, [ c_\sigma(a)-c_\tau(a) ]
  \left[2P_0-(P_\sigma+P_\tau)\right],
\end{equation}
where $P_0$ is the plaquette expectation value on a symmetric
($T=0$) lattice, and the \textit{anisotropy coefficients}
$c_{\sigma,\tau}(a)$ are defined by:
\begin{equation}\label{e:cst}
    c_{\sigma,\tau}(a)\equiv
    \left(\frac{\partial g^{-2}_{\sigma,\tau}}
    {\partial \xi}\right)_{\xi=1}.
\end{equation}
\begin{figure}
\centerline{\psfig{file=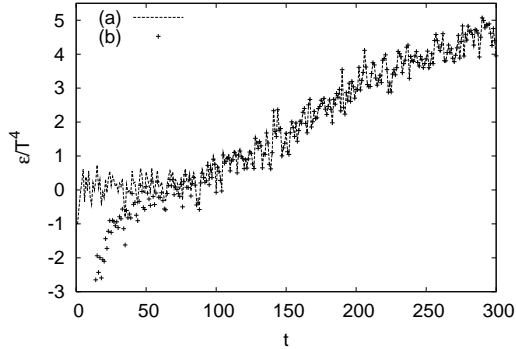,width=7cm}} \vspace*{8pt}
\caption{SU(3) gluonic energy density: (a) with $P_0$ calculated
from the time series after the quench and (b) using equilibrium
values for $P_0$.} \label{fig_su3subtract}
\end{figure}
They are related to the QCD $\beta$-function and can be calculated
using Pade fits of Ref.~\refcite{BoEn96}. As was shown in
Ref.~\refcite{BBV01} this procedure can be successfully used for
the quench. To normalize to zero temperature, plaquette values
from the symmetric $N_{\tau}=N_{\sigma}$ lattice are needed in
Eq.~(\ref{e:eminus3p}). As one stays within the confined phase on
the symmetric lattice its equilibration after the quench is fast.
Therefore it is enough to use equilibrium values of $P_0$ (at
final $\beta_f=5.92$) after the quench. This is illustrated
in Fig.~\ref{fig_su3subtract}.

\begin{figure}
\centerline{\psfig{file=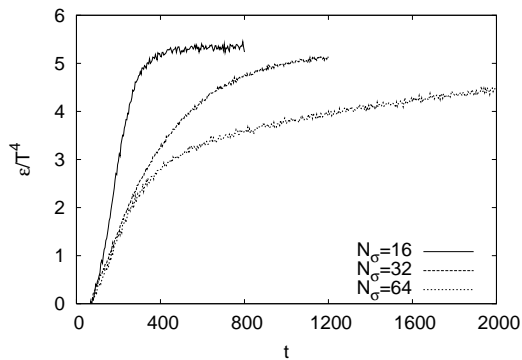,width=7cm}} \vspace*{8pt}
\caption{The time evolution of SU(3) gluonic energy density on
$4\times N_\sigma^3$ lattices.} \label{fig_geds}
\end{figure}

In Fig.~\ref{fig_geds} we report how the time needed for equilibration
increases with lattice size. We compare the evolution of the gluonic
energy density on lattices $4\times 16^3$, $4\times 32^3$ and
$4\times 64^3$. A slowing down of the equilibration is found, which
appears to be related to the divergence of the time needed to reach
the structure factor maxima.

\begin{figure}
\centerline{\psfig{file=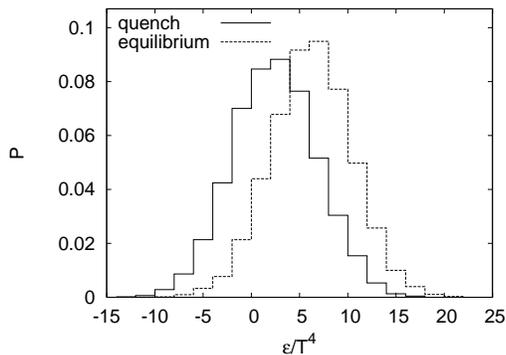,width=7cm}}
\vspace*{8pt}
\caption{SU(3) gluonic energy density $P(\epsilon)$ histograms
on a $4\times 16^3$ lattice: (a) with competing vacuum domains present
and (b) after reaching equilibrium.} \label{fig_su3ged_2hists}
\end{figure}

Finally, we compare in Fig.~\ref{fig_su3ged_2hists} for the
$N_{\sigma}=16$ lattice the gluonic energy distribution in equilibrium
at $\beta=5.92$ with the one obtained after 148 time steps. We find
a shift towards lower gluonic energies and the width of the probability
density is slightly broader for the time evolution after the quench
than in equilibrium.
One also has to take into account that the geometry of relativistic
heavy ion experiments experiments is reasonably approximated by
$N_{\tau}/N_{\sigma}={\rm const}$, $N_{\sigma}\to\infty$, rather
than by $N_{\tau}={\rm const}$, $N_{\sigma}\to\infty$.

\section{Summary and Conclusions\label{sec_summary}}

Using the evolution of structure function modes, we identify
spinodal decomposition as the transition scenario for quench from
the disordered into the ordered phase of SU(3) lattice gauge
theory and the analogue Potts model. We observe an early time
development of structure factors $S_n(t)$, which is in over-all
agreement with the exponential growth predicted by the linear
theory of spinodal decomposition for $|\vec{p}|<p_c$. From our
data the critical mode $p_c$ is estimated. Using  phenomenological
arguments~\cite{MiOg02}, $p_c$ is used to determine the Debye
screening mass $m_D$ at the final temperature $T$.

With increasing lattice size $N_{\sigma}$ the time to reach the
structure factor maxima diverges.  Relying on a study of
Fortuin-Kasteleyn clusters in Potts models \cite{BMV}, we assume
that the reason for the slowing down is due to competing vacuum
domains. For SU(3) gauge theory these are domains of distinct
$Z_3$ trialities. They may be the relevant vacuum configurations
after the heating quench in relativistic heavy ion collision
experiments. We have initiated a study of the gluonic energy and
pressure densities on such configurations. Our data are consistent
with a divergence of the equilibration time of the gluonic energy
density, similarly to the one observed for the structure factor
maxima.

All our results rely on using a dissipative, non-relativistic time
evolution, believed to be in the Glauber universality class. The
hope is that the thus created non-equilibrium configurations may
exhibit some features, which are in any dynamics typical for the
state of the system after the quench. This hope could get more
credible by studying a Minkowskian time evolution of Polyakov
loops and finding similar features. Such a study appears to be
possible \cite{Du01} within a relativistic Polyakov loop model
which was introduced by Pisarski \cite{DuPi01}.

\section*{Acknowledgments}

This work was in part supported by the US Department of Energy under
contract DE-FG02-97ER41022. The simulations were performed on PC
clusters at FSU.

\end{document}